\documentclass[sn-mathphys,Numbered,iicol]{sn-jnl}% Math and Physical Sciences Reference Style
\usepackage{graphicx}%
\usepackage{multirow}%
\usepackage{amsmath,amssymb,amsfonts}%
\usepackage{amsthm}%
\usepackage{mathrsfs}%
\usepackage[title]{appendix}%
\usepackage{xcolor}%
\usepackage{textcomp}%
\usepackage{manyfoot}%
\usepackage{booktabs}%
\usepackage{algorithm}%
\usepackage{algorithmicx}%
\usepackage{algpseudocode}%
\usepackage{listings}%

\begin{document}

\title[Can IKEA effect promote empathy for agents?]{Can IKEA effect promote empathy for agents?}

%%=============================================================%%
%% Prefix	-> \pfx{Dr}
%% GivenName	-> \fnm{Joergen W.}
%% Particle	-> \spfx{van der} -> surname prefix
%% FamilyName	-> \sur{Ploeg}
%% Suffix	-> \sfx{IV}
%% NatureName	-> \tanm{Poet Laureate} -> Title after name
%% Degrees	-> \dgr{MSc, PhD}
%% \author*[1,2]{\pfx{Dr} \fnm{Joergen W.} \spfx{van der} \sur{Ploeg} \sfx{IV} \tanm{Poet Laureate} 
%%                 \dgr{MSc, PhD}}\email{iauthor@gmail.com}
%%=============================================================%%

\author*[1,2]{\fnm{Takahiro} \sur{Tsumura}}\email{takahiro-gs@nii.ac.jp}

\author[2,1]{\fnm{Seiji} \sur{Yamada}}%\email{seiji@nii.ac.jp}

\affil*[1]{\orgdiv{Department of Informatics}, \orgname{The Graduate University for Advanced Studies, SOKENDAI}, \orgaddress{\street{Shonan Village, Hayama}, \city{Kanagawa}, \postcode{240-0193}, \country{Japan}}}

\affil[2]{\orgname{National Institute of Informatics}, \orgaddress{\street{2-1-2 Hitotsubashi, Chiyoda-ku}, \city{Tokyo}, \postcode{101-8430}, \country{Japan}}}

%%==================================%%
%% sample for unstructured abstract %%
%%==================================%%

\abstract{
Cooperative relationships between humans and agents are becoming more important for the social coexistence of anthropomorphic agents, including virtual agents and robots. 
One way to improve the relationship between humans and agents is for humans to empathize with the agents. 
Empathy can help humans become more accepting of agents. 
In this study, we focus on the IKEA effect in creating agents and investigate human empathy toward agents through relationships with others in the same space. 
For this reason, this study used a robot assembly task in which two participants cooperatively build the same robot or individually their own robot. 
We conducted experiments to examine the relationship between participants, the IKEA effect in creating an agent, and the influence of the empathy object on human empathy. 
The results showed that the IKEA effect promoted empathy toward the agent regardless of the relationship between participants. 
On the other hand, there was no significant difference in empathy from one participant to another before and after the task. 
These results indicate that regardless of the relationship between participants in the same space, the creation of an agent can promote empathy toward the agent.
}


%%\pacs[JEL Classification]{D8, H51}

%%\pacs[MSC Classification]{35A01, 65L10, 65L12, 65L20, 65L70}

\maketitle

\section{Introduction}
Many intelligent artifacts, such as ChatGPT, generative AI, and social robots, are widely used in human society. 
Humans live in societies and use a variety of tools, and artifacts are sometimes treated as if they were human. Humans are known to tend to treat artifacts as if they were humans in the media equation~\cite{Reeves96}. 
However, these objects are so prevalent that humans may not consider some of them to be human-like. 
A possible solution to the problem with AI and robots is to focus on their relationship to artifacts.
\\ \indent
Previous AI issues have focused on the trustworthy and ethical use of AI technology. 
Ryan~\cite{Ryan20} focused on trust and discussed the issue of AI ethics and people anthropomorphizing AI. 
It was determined that even complex machines that use AI should not be considered trustworthy. 
Instead, he suggested, we need to ensure that organizations that use AI and the individuals within those organizations are trustworthy. 
Belk~\cite{Belk20} filled a gap in recent research on AI and robotics in services, expanding the view of the service context in which robotics and AI are involved and providing important implications for public policy and service technology applications. 
AI ethics was also discussed in detail in terms of applied ethics in a study by Hallamaa and
Kalliokoski~\cite{Hallamaa22}. 
Wirtz and Pitardi~\cite{Wirtz23} noted that the shift to robotics and AI-enabled services will simultaneously improve customer experience, service quality, and productivity, but they mentioned the ethical, fairness, and privacy risks for customers and society. 
They therefore discussed the impact of the service revolution on service firms, their marketing, and their customers and offered possibilities for future research. 
Kumar et al.~\cite{Kumar23} noted that new technological advances made possible with the use of AI in medicine have not only raised concerns about public trust and ethics but have also generated much debate about its integration into medicine. 
They reviewed current research investigating how to apply AI technologies to create smart predictive maintenance.
\\ \indent
Similar to our trust in AI, we often empathize with artifacts.
There are several previous studies that have investigated the relationship between trust and empathy. 
Spitale et al.~\cite{Spitale22} investigated the amount of empathy elicited by a social assistance robot storyteller and the factors that influence the user's perception of that robot. 
As a result, the robot narrator elicited more empathy when the object of the story's empathy matched that of the narrator. 
Johanson et al.~\cite{Johanson23} investigated whether the use of verbal empathic statements and nods from a robot during video-recorded interactions between a healthcare robot and patient would improve participant trust and satisfaction. 
Results showed that the use of empathic statements by the healthcare robot significantly increased participants' empathy, trust, and satisfaction with the robot and reduced their distrust of the robot. 
\\ \indent
Thus, we empathize with cleaning robots, pet robots, anthropomorphic agents, etc. that provide services for online shopping and help desks. 
These are already in use in society and coexist with humans. 
The appearance of these agents also varies depending on the application and the environment in which they are used. 
However, some humans cannot accept this type of agent~\cite{Nomura08,Nomura16}. 
\\ \indent
Mori~\cite{Mori70} proposed an uncanny valley of human emotional responses to robots, in which robots become more likeable and empathetic as they are made more human-like in their appearance and behavior, but at some point suddenly turn into strong dislike.
As ever-developing technology acquires human-like capabilities, the situation becomes similar to the uncanny valley, leading to negative behavior.
With respect to the uncanny valley hypothesis, Thepsoonthorn et al.~\cite{Thepsoonthorn21} explored the uncanny valley in terms of the robot's nonverbal behavior.
The results showed a two-correlation coefficient between the results of human-like and affinity ratings. A curve similar to the uncanny valley was found.
A curve similar to the uncanny valley was suggested for nonverbal behavior by them, indicating that nonverbal behavior that is more human-like is thought negatively by people.
\\ \indent
Mahmud et al.~\cite{Mahmud22} also conducted a literature review of algorithm aversion, which is a negative behavior or attitude toward the workings of algorithms compared to the workings of humans, and has gained attention through the use of ChatGPT and generative AI.
In particular, Filiz et al.~\cite{Filiz23} examined the relationship between the outcome of a decision-making situation and the frequency of algorithm aversion and showed that the more serious the outcome of a decision, the more frequently algorithm aversion occurred.
Especially in the case of very important decisions, algorithmic aversion leads to a reduced probability of success, which is the tragedy of algorithmic aversion.
For causes that lead to the negative behavior described above, it is an important issue to establish a coexistence relationship with AI technologies and anthropomorphic agents that will be used in the future society.
\\ \indent
The following research focuses on promoting human empathy toward agents. 
Tsumura and Yamada~\cite{Tsumura22} previously focused on ``(human) empathy" as an attribute that agents should possess in order to be accepted by human society and aimed to create empathy agents. 
Tsumura and Yamada~\cite{Tsumura23-1} focused on self-disclosure from agents to humans in order to enhance human empathy toward anthropomorphic agents, and they experimentally investigated the potential for self-disclosure by agents to promote human empathy. 
Tsumura and Yamada~\cite{Tsumura23-2} also focused on tasks in which humans and agents engage in a variety of interactions, and they investigated the properties of agents that have a significant impact on human empathy toward them. 
In this study, we focused on how humans can improve their relationships with agents. 
Therefore, similar to these studies, we focused on human empathy toward agents as one way to improve the relationship between humans and agents. 
Empathy makes humans more likely to act positively toward an agent and more likely to accept the agent. 
Factors that generate empathy, including linguistic information, nonverbal information, situation, and relationship, have been studied in various ways. 
Shaffer et al.~\cite{Shaffer19} showed that narrative writing intervention increased participants' empathy and perspective-taking, evoked more positive attitudes toward women who smoke during pregnancy, and increased external attributions for the behavior of these women.
\\ \indent
Previous studies have shown that assembling artifacts, known as the IKEA effect~\cite{Norton12}, is important for increasing the value of objects. 
However, no study has used the IKEA effect in investigating empathy for agents. 
We hypothesize that the IKEA effect is necessary for anthropomorphic agents to have a better relationship with humans. 
In recent years, as agents have been increasingly utilized in human society, the establishment of relationships between humans and agents has become increasingly important. 
Just as humans enhance the value of artifacts through the IKEA effect, it makes sense for agents to use the IKEA effect to deepen their relationships with humans.
\\ \indent
In terms of the media equation, it has also been suggested that people feel the same emotions toward artifacts as they do toward people, but it is not clear how much empathy for artifacts differs from empathy for people. 
Therefore, in this study, we investigate two objects of empathy toward agents and humans. 
The empathy addressed in this study concerns empathy characteristics toward people and agents.

\section{Related work}
\subsection{Definition of empathy}
We consider empathy to be a significant element in being accepted by humans as a member of society. 
For humans to get along with each other, it is important that they empathize with the other~\cite{Gaesser13,Klimecki16}. 
\\ \indent
Empathy and the effects that it has on others have been a focus of research in the field of psychology. 
Omdahl~\cite{Omdahl95} roughly classifies empathy into three types: (1) affective empathy, which is an emotional response to the emotional state of others, (2) cognitive understanding of the emotional state of others, which is defined as cognitive empathy, and (3) empathy including the above two. 
Preston and de Waal~\cite{Preston02} suggested that at the heart of the empathic response was a mechanism that allowed the observer to access the subjective emotional state of the target. 
They defined the perception-action model (PAM) and unified the different perspectives in empathy. 
They defined empathy as three types: (a) sharing or being influenced by the emotional state of others, (b) assessing the reasons for the emotional state, and (c) having the ability to identify and incorporate other perspectives. 
Arbel et al.~\cite{Arbel21} addressed the gap by focusing on adaptive empathy, defined as the ability to learn and adjust one's empathic response based on feedback. 
They provided a laboratory-based model for studying adaptive empathy and demonstrated the potential contribution of learning theory to increasing our understanding of the dynamic nature of empathy.
\\ \indent
Although we focus on the positive effects of empathy to improve society's acceptance of AI agents in this study, empathy has been discussed from various aspects including negative effects in psychological literature. 
Bloom~\cite{Bloom16} tried to introduce a neutral aspect of empathy by introducing not only positive influences but also negative ones. 
He claimed that it is possible for empathy to act as a moral guide that leads humans to irrational decision-making and relationships to violence and anger. 
Also, he claimed that we can overcome this problem by using conscious, deliberative reasoning and altruistic approaches. 
\\ \indent
Various questionnaires are used as a measure of empathy, but we examined two famous ones. 
The Interpersonal Reactivity Index (IRI), also used in the field of psychology, is used to investigate the characteristics of empathy~\cite{Davis80}. 
Baron-Cohen and Wheelwright~\cite{Baron-Cohen04} reported a new self-report questionnaire, the Empathy Quotient (EQ), for use with adults of normal intelligence. 
Lawrence et al.~\cite{Lawrence04} investigated the reliability and validity of the EQ and determined its factor structure. 
The experimental results showed a moderate association between the EQ subscale and the IRI subscale.
\\ \indent
Regarding questionnaires about empathy, IRI and EQ are widely used in the field of psychology. 
In particular, examples of research on empathy using only the IRI are widely seen in psychology and the HRI and HAI fields, such as the studies by Shaffer et al.~\cite{Shaffer19} and Tsumura and Yamada~\cite{Tsumura23-1,Tsumura23-2}. 
Since the results of previous studies have shown that the EQ scale is related to the IRI scale, we used the IRI questionnaire, which has fewer questions and allows for the four characteristics of empathy to be investigated. 
We also used the widely used IRI for comparison with previous and future research on empathy. 
We also focused on investigating the impact of each characteristic of empathy. 
The use of the IRI was appropriate for investigating this as based on previous studies.

\subsection{Empathy in engineering}
Empathy has also been studied in the field of engineering, particularly in the context of virtual reality. 
vanLoon et al.~\cite{vanLoon18} investigated whether the effects of VR perspective-taking could be driven by increased empathy and extended to real-stakes behavioral games. 
They succeeded in increasing the tendency of participants to take the other person's point of view, but only if it was that of the same person participants assumed in the virtual reality simulation. 
Herrera et al.~\cite{Herrera18} compared the short- and long-term effects of traditional and VR viewpoint acquisition tasks. 
They also conducted experiments investigating the role of technological immersion with respect to different types of intermediaries. 
\\ \indent
Tassinari et al.~\cite{Tassinari22} developed an inter-participant design to investigate how VR can be used to create positive inter-group contact with members of prejudiced out-groups (ethnic minorities) and presented results of the impact of inter-group contact in VR on empathy. 
Crone and Kallen~\cite{Crone22} utilized online platforms and immersive virtual reality to examine the role of virtual perspective-taking on binary gender. 
They found that virtual reality-based perspective-taking may have a greater impact on acute behavioral modulation of gender bias compared with online because it immerses participants in the experience of temporarily becoming another.
\\ \indent
Empathy is also attracting attention in the realm of product design. 
Bennett and Rosner~\cite{Bennett19} investigated a human-centered design process (promise of empathy) in which designers try to understand the target user with the aim of informing technology development. 
Rahmanti et al.~\cite{Rahmanti22} designed a chatbot with artificial empathic motivational support for dieting called ``SlimMe'' and investigated how people responded to the diet bot. 
They proposed a text-based emotional analysis that simulates artificial empathic responses to enable the bot to recognize users' emotions.
\\ \indent
Drouet et al.~\cite{Drouet22} shared the preparatory stages of developing an empathy scale for service design, as other stakeholders had overlooked the need to foster empathy for users to provide high-quality user-centered services and products. 
Al-Farisi et al.~\cite{Al-Farisi22} applied two designs of anthropomorphic design cues (ADCs) (verbal and nonverbal designs) to chatbots and compared the empathy levels of the two chatbots with and without ADCs. 
The ADC had a significantly positive effect on increasing the chatbot's level of empathy.

\subsection{Empathy in human-robot/agent interaction}
Studies on human empathy in the human-robot interaction (HRI) field have explored the ways in which humans empathize with artificial objects. 
Yamada and Komatsu~\cite{Yamada06} proposed a design policy called ``SE2PM: simple expression to primitive mind" using a mobile robot that expresses its mind with beeps based on Mindstorms and a pet robot, AIBO, that expresses its complex behavior with mental expressions, and using this design policy, they implemented the robot's mental expressions at full scale and examined their effectiveness. 
SE2PM suggested that intuitive and simple expressions such as beeps of simple robots (e.g., mobile robots) are more effective in conveying their primitive minds to humans than complex behaviors of complex robots (e.g., pet robots such as dogs).
\\ \indent
On the basis of the concept of cognitive developmental robotics, Asada~\cite{Asada15} proposed ``affective developmental robotics" as a way to produce more authentic artificial empathy. 
Artificial empathy here refers to AI systems (e.g., companion robots and virtual agents) that can detect emotions and respond empathetically. 
The design of artificial empathy is one of the most essential components of social robotics, and empathetic interaction with the public is necessary to introduce robots into society.
\\ \indent
Fraune~\cite{Fraune20} examined how people behave morally and perceive players according to their group membership (in-group, out-group), agent type (human, robot), and robot anthropomorphism (anthropomorphic, mechanized). 
Their results showed that the pattern of reactions to humans was more favorable for anthropomorphic robots than for mechanistic robots. 
Park and Whang\cite{Park22} conducted a systematic review of the literature on empathy in interpersonal, virtual agent, and social robot research, using inclusion criteria for analyzing empirical studies in peer-reviewed journals, conference proceedings, or articles. 
They suggested key factors such as domain dependence, multimodality, and empathic coordination to consider when designing, engineering, and studying empathic social robots.
\\ \indent
As for empathy research in the field of human-agent interaction (HAI), Leite et al.~\cite{Leite14} conducted a long-term study in an elementary school where they presented and evaluated an empathy model for social robots aimed at interactions with children that occur over a long period of time.
They measured children's perceptions of social presence, engagement, and social support and found that the empathy model developed had a positive impact on the long-term interaction between the child and the robot. 
Chen and Wang~\cite{Chen19} hypothesized that empathy and anti-empathy were closely related to a creature's inertial impression of coexistence and competition within a group and established a unified model of empathy and anti-empathy. 
They also presented the Adaptive Empathetic Learner (AEL), an agent training method that enables evaluation and learning procedures for emotional utilities in a multi-agent system. 
\\ \indent
Perugia et al.~\cite{Perugia20} investigated which personality and empathy traits were related to facial mimicry between humans and artificial agents. 
They focused on the humanness and embodiment of agents and the influence that these have on human facial mimicry. 
Their findings showed that mimicry was affected by the embodiment that an agent has, but not by its humanness. 
It was also correlated with both individual traits indicating sociability and empathy and traits favoring emotion recognition. 
Parmar et al.~\cite{Parmar22} investigated the systematic manipulation of animation quality, voice quality, rendering style, and simulated empathy and their impact on virtual agents' perceptions in terms of naturalness, engagement, trust, credibility, and persuasion in health states. 
Tsumura and Yamada~\cite{Tsumura23-3} experimentally examined the hypothesis that agent reactions and human preferences affect human empathy and acceptance of agent mistakes. 
The results showed that agent reactions and human preferences do not affect empathy toward the agent, but do allow the agent to make mistakes. 
It was also shown that empathy for an agent decreases when the agent makes mistakes in a task.
\\ \indent
To clarify the empathy between agents/robots and humans, Paiva represented the empathy and behavior of empathetic agents (called empathy agents in HAI and HRI studies) in two different ways: targeting empathy and empathizing with observers Paiva et al.~\cite{Paiva17}. 
In our study, following Paiva's proposed definition of an empathic agent, we consider the agent as an object of empathy and examine how the empathy and empathic responses of human participants are affected.

\subsection{IKEA effect and human relationship}
Sun and Sundar~\cite{Sun16} investigated whether assembling a robot enhances the quality of interaction with the robot and whether it matters whether the robot is a practical tool or a socially interacting entity. 
Results showed that when participants set up the robot themselves, they tended to have more positive evaluations of both the robot and the interaction process, with effects positively mediated by a sense of ownership and accomplishment and negatively mediated by perceived process costs. 
Marsh et al.~\cite{Marsh18} elucidated the developmental stages of the IKEA effect and showed that the bias emerges at age 5, an important developmental milestone in the formation of self-concept. 
They also assessed the role of effort and found that the IKEA effect was not moderated by the amount of effort expended on the task in 5- to 6-year-olds. 
They further examined whether feelings of ownership moderated the IKEA effect and found that feelings of ownership alone did not explain why children valued their work more highly. 
Wald et al.~\cite{Wald21} investigated the potential of active user-based chatbot customization for the development of trust in chatbots. 
While customization had no direct impact on trust, anthropomorphism was identified as an important mediating factor.
\\ \indent
Aeschlimann et al.~\cite{Aeschlimann20} investigated communication patterns and prosocial outcomes in interactions with voice assistants. 
Results showed that children did not have the same expectations of voice assistants as they did of humans, and that human-to-human and human-to-computer cooperation differed. 
Pauw et al.~\cite{Pauw22} examined the socio-emotional benefits of talking to virtual humans and whether these benefits are moderated by the type of support provided. 
To examine the range of potential effects, they compared the two main types of support (emotional and cognitive) across two emotions (anger and worry). 
Results revealed that participants felt better after speaking with a virtual human, the intensity of the targeted emotion decreased, and their overall emotional state improved.
\\ \indent
Spaccatini et al.~\cite{Spaccatini23} experimentally investigated whether the type of mind attributed to anthropomorphic social robots has a complementary effect on empathy for people in need. 
The results showed that anthropomorphism promoted the attribution of subjectivity (anthropomorphic appearance and interaction via chatbot) and experience (anthropomorphic appearance only). 
Jorge et al.~\cite{Jorge23} investigated how automation failures in human-automation collaborative scenarios affect human confidence in automation as well as human confidence in automation. 
The results showed that automation failures negatively affect human trust and their trust and liking for automation.

\section{Materials and methods}
\subsection{Ethics statement}
The protocol was approved by the ethics committee of the National Institute of Informatics (13, April, 2020, No. 1). 
All studies were carried out in accordance with the recommendations of the Ethical Guidelines for Medical and Health Research Involving Human Subjects provided by the Ministry of Education, Culture, Sports, Science and Technology and Ministry of Health, Labour and Welfare in Japan. Written informed consent was provided by choosing one option on an online form: ``I am indicating that I have read the information in the instructions for participating in this research. 
I consent to participate in this research." All participants gave informed consent. 
After that, they were debriefed about the experimental procedures. 

\subsection{Hypotheses}
The purpose of this study was to determine whether assembling an agent would generate the IKEA effect and promote human empathy toward the agent. 
We will also investigate whether the relationship between participants during the task affects their empathy toward the agent they assemble.
For these purposes, we used LEGO Mindstorms EV3 to investigate the empathy that participants have for the agent and another participant before and after the task. 
For this study, we formulated the following three hypotheses.
\vspace{1mm}
\begin{enumerate}
\item[\textbf{H1}:] When an agent is assembled, empathy for the agent is promoted by the IKEA effect.
\item[\textbf{H2}:] When participants cooperate, empathy for the other participant is promoted.
\item[\textbf{H3}:] When participants create agents individually, empathy for the agents is promoted.
\end{enumerate}
\vspace{1mm}

The reason for Hypothesis 1 is to investigate whether the IKEA effect influences empathy as one of the factors that make people judge something to be more valuable than others. 
Hypotheses 2 and 3 were formulated because it is well known that empathy for a partner increases after several people perform a task together~\cite{Sun16, Fraune20, Pauw22}, and it is possible that the evaluation of the object of empathy will change depending on the relationship of the task at this time.
\\ \indent
We designed a mixed design experiment using three factors: the IKEA effect, participants' relationships, and the empathy object. 
The number of levels for each factor was two for the IKEA effect (before and after), two for the participants' relationships (cooperative and individual), and two for the empathy object (agent and another participant). 
Participant relationships were inter-participant factors, while the IKEA effect and empathy object were intra-participant factors. 
The dependent variable was the empathy felt by the participant.

\subsection{LEGO Mindstorms EV3}
\begin{figure}[tbp]
\centering
\includegraphics[scale=0.36]{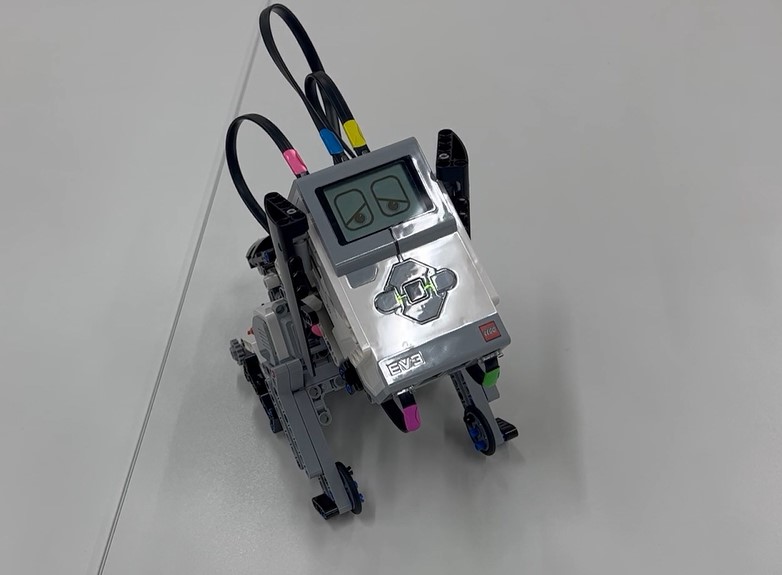}
\caption{LEGO agent.}
\label{fig2}
\end{figure}

\begin{figure}[tbp]
\centering
\includegraphics[scale=0.28]{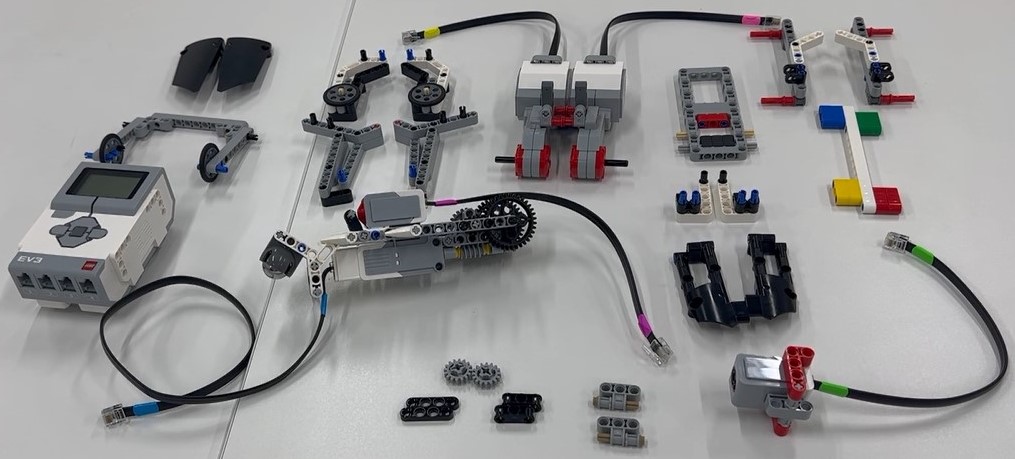}
\caption{Parts before assembly.}
\label{fig3}
\end{figure}

The LEGO Mindstorms EV3 (``LEGO Agent") used in this study is an assemblable robot. 
The participants assembled a dog-shaped robot during the experiment, one of the possible forms of LEGO Mindstorms EV3.
Figure~\ref{fig2} shows the finished robot assembled in the task.
\\ \indent
Although its behavior can be controlled by programming, programming was not required for this study, so participants only had to confirm that the sample program worked successfully after the assembly was completed.
In this experiment, to simplify the assembly process, we assembled the small parts of the robot in advance, and the finished product shown in Figure~\ref{fig3} was assembled from its disassembled state. 
Information on this LEGO agent is summarized below.
\begin{description}
\item[\textbf{Model number}:] EVR45544S
\item[\textbf{Weight}:] 791 g
\item[\textbf{Height}:] 17.2 cm
\item[\textbf{Average assembly time}:] 45 minutes
\end{description}

\subsection{Experimental procedure}
\begin{figure*}[tbp]
\centering
\includegraphics[scale=0.5]{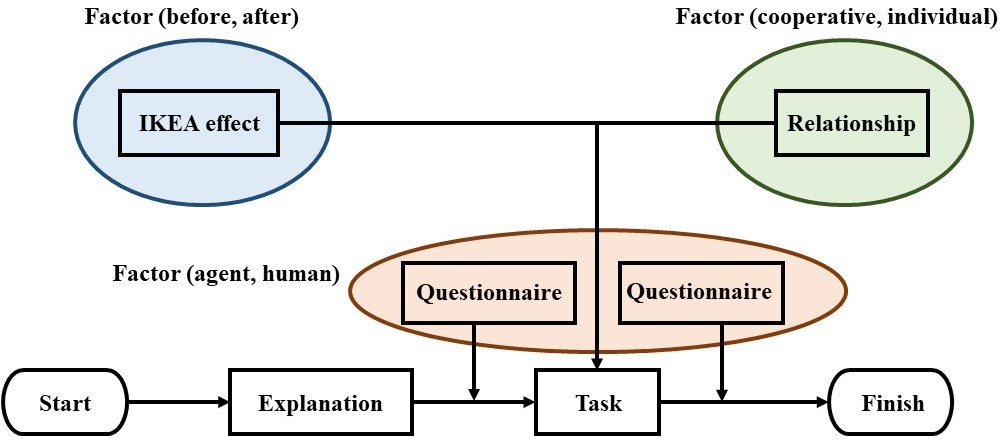}
\vspace{1mm}
\caption{Process flow of experiment.}
\label{fig}
\end{figure*}

\begin{figure*}[tbp]
\centering
\includegraphics[scale=0.4]{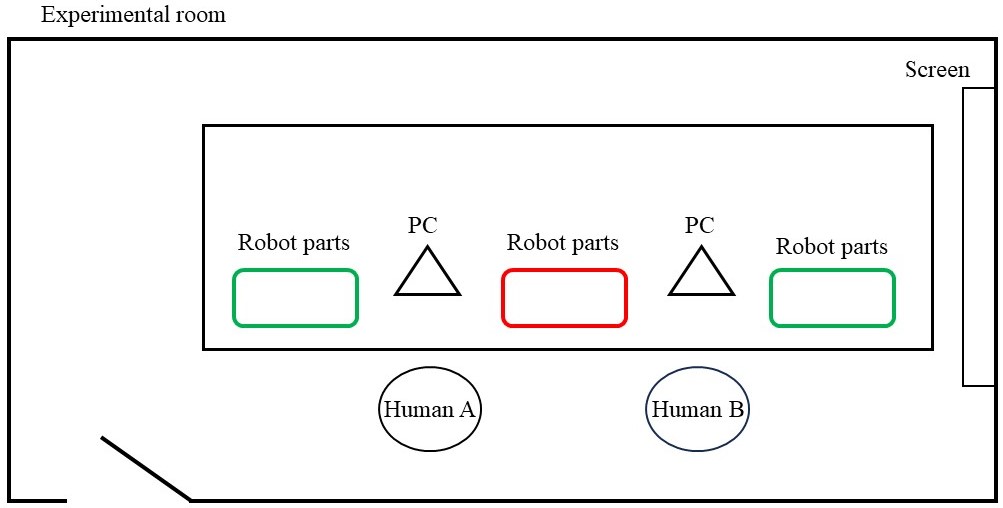}
\vspace{1mm}
\caption{Image of experimental environment.}
\label{fig1}
\end{figure*}

This experiment was conducted at the National Institute of Informatics with participants recruited through a recruitment website. 
The process flow is shown in Figure~\ref{fig}. 
Since our main goal was to promote human empathy for anthropomorphic agents, we administered a questionnaire to measure participants' empathy for the agent and another participant before they performed the task.
Figure~\ref{fig1} is an image of the experimental environment. 
The following explanation is based on this figure.
\\ \indent
Before the pre-task questionnaire, we showed a video of the finished agent, which was created by us during the explanation phase, in action on screen. 
Participants sat in seats A and B and worked on assembling the robot. 
The robot parts were placed in different locations depending on the experimental conditions: in the case of cooperative relationships, the parts were placed in the red locations; in the case of individual relationships, the parts were placed in the green locations.
The PC in front of the participant was prepared for the survey and for viewing the assembly video. 
Participants were allowed to talk to each other during the creation of the agent. 
The reason for this setting was that this experiment could not assume that participants would be prohibited from being in the same space and talking when creating the agents.
\\ \indent
The task was for participants to assemble a dog-shaped LEGO agent; if the participants were in a cooperative relationship, they assembled one agent; if they were in individual relationships, each participant assembled an agent.
The entire experiment took approximately one hour to complete. 
The participants were left alone in the room during the assembly process.
The purpose of this experiment was to investigate the IKEA effect, which was a factor in this experiment, and the influence of the participants' relationship on the empathy object.
\\ \indent
Participants were given a few minutes after the completion of the LEGO agent to check its simple operation and time to operate it. 
This was a time for the participants to recognize that the agent they had assembled was actually moving and had been completed successfully.
The behavior of the agent was similar to the behavior video shown during the explanation.
The specific behavior can be seen in the video attached to the supplemental materials. 
After the task was completed, the same questionnaire was administered as before the task.
After completing the questionnaire, participants spontaneously responded to open-ended questions.

\subsection{Participants}
Through the recruitment website, participants were paid 1,500 yen (= 10.27 dollars). 
The number of participants was 40. 
To ensure that the number of participants was sufficient for the statistical analysis of the three-factor mixed design, we used G*power, a software program that can check the appropriate number of participants for statistical analysis~\cite{Faul07}. 
The following conditions were used in the study: 24 participants were available for the analysis; thus, 40 participants were found to be a sufficient number for the analysis (Conditions: F tests, ANOVA: Repeated measures, within-between interaction, Effect size f: 0.25, $\alpha$ error prob: 0.05, Power (1-$\beta$ error prob): 0.8, Number of groups: 2, Number of measurements: 4, Corr among rep measures: 0.5, Nonspheric correction $\epsilon$: 1).
\\ \indent
Cronbach's $\alpha$ coefficient was then applied to the 40 participants to determine the reliability of their survey responses and found to be between 0.723 and 0.890 in all conditions. 
Twenty participants in each condition were included in the analysis. 
The mean age was 26.38 years (standard deviation 7.693), with a minimum of 18 years and a maximum of 52 years. 
Gender was 25 males and 15 females.

\subsection{Questionnaire}
Participants answered questionnaires before and after the task. 
This was a 12-item questionnaire modified from the IRI.
As the IRI was designed to investigate human empathy characteristics, we modified it to investigate empathy toward an empathy agent and another participant. 
The part of the description that says ``character" changed between LEGO agent or other participant. 
The same questionnaire was applied both before and after the task and was administered on a 5-point Likert scale (1: not applicable, 5: applicable), as shown in Table~\ref{table1}. 
Q4, Q9, and Q10 are inverted items, so the scores were reversed when analyzing them.
\\ \indent
Q1 to Q6 examine affective empathy, and Q7 to Q12 examine cognitive empathy. 
There was one additional item for the questionnaire administered after the task (BeQ in the table) to examine the empathic response of the participants. 

\renewcommand{\arraystretch}{1.1}
\begin{table*}[tbp] 
    \caption{Summary of questionnaire}
    \vspace{1mm}
    \centering
    \scalebox{0.84}{
    \begin{tabular}{l}\hline 
        \textbf{Affective empathy}\\ \hline
        \textbf{Personal distress}\\
        Q1: If an emergency happens to the character, you would be anxious and restless.\\
        Q2: If the character is emotionally disturbed, you would not know what to do.\\
        Q3: If you see the character in need of immediate help, you would be confused and would not know what to do.\\
        \textbf{Empathic concern}\\
        Q4: If you see the character in trouble, you would not feel sorry for that character.\\
        Q5: If you see the character being taken advantage of by others, you would feel like you want to protect that character.\\
        Q6: The character's story and the events that have taken place move you strongly.\\\hline
        \textbf{Cognitive empathy}\\ \hline
        \textbf{Perspective taking}\\
        Q7: You look at both the character's position and the human position.\\
        Q8: If you were trying to get to know the character better, you would imagine how that character sees things.\\
        Q9: When you think you're right, you don't listen to what the character has to say.\\
        \textbf{Fantasy scale}\\
        Q10: You are objective without being drawn into the character's story or the events taken place.\\
        Q11: You imagine how you would feel if the events that happened to the character happened to you.\\
        Q12: You get deep into the feelings of the character.\\\hline
    \end{tabular}}
    \label{table1}
\end{table*}
\renewcommand{\arraystretch}{1.0}

\subsection{IKEA effect}
As mentioned, to generate the IKEA effect in this study, a LEGO Mindstorms EV3 dog-shaped robot was assembled. 
Figure~\ref{fig2} shows the finished robot assembled in the task. 
The IKEA effect in LEGO-based assembly tasks was already investigated in a study by Norton et al.~\cite{Norton12}. 
Therefore, we considered the IKEA effect to have occurred before and after the LEGO agent was assembled in this experiment.
To investigate changes in empathy toward the agent before and after the IKEA effect, we decided to show a video of the finished robot before the assembly task.
\\ \indent
In this experiment, to simplify the assembly process, we assembled the small parts of the robot in advance, and the finished product shown in Figure~\ref{fig3} was assembled from its disassembled state. 
For the assembly task, we created a video of the assembly process, and participants watched the video while assembling the LEGO agent.
\\ \indent
The assembly process did not vary greatly between cooperative and individual work, taking an average of 45 minutes. 
The specific assembly procedure is illustrated in Figure~\ref{figa} below. 
First, the fuselage is assembled (No. 2). Next, the front legs are attached to the body (No. 3). 
Then, the back legs are attached to the body (No. 4).
After that, the neck is attached to the body (No. 5). 
Finally, the head is attached to the neck (No. 6).

\begin{figure*}[tbp]
\centering
\includegraphics[scale=0.4]{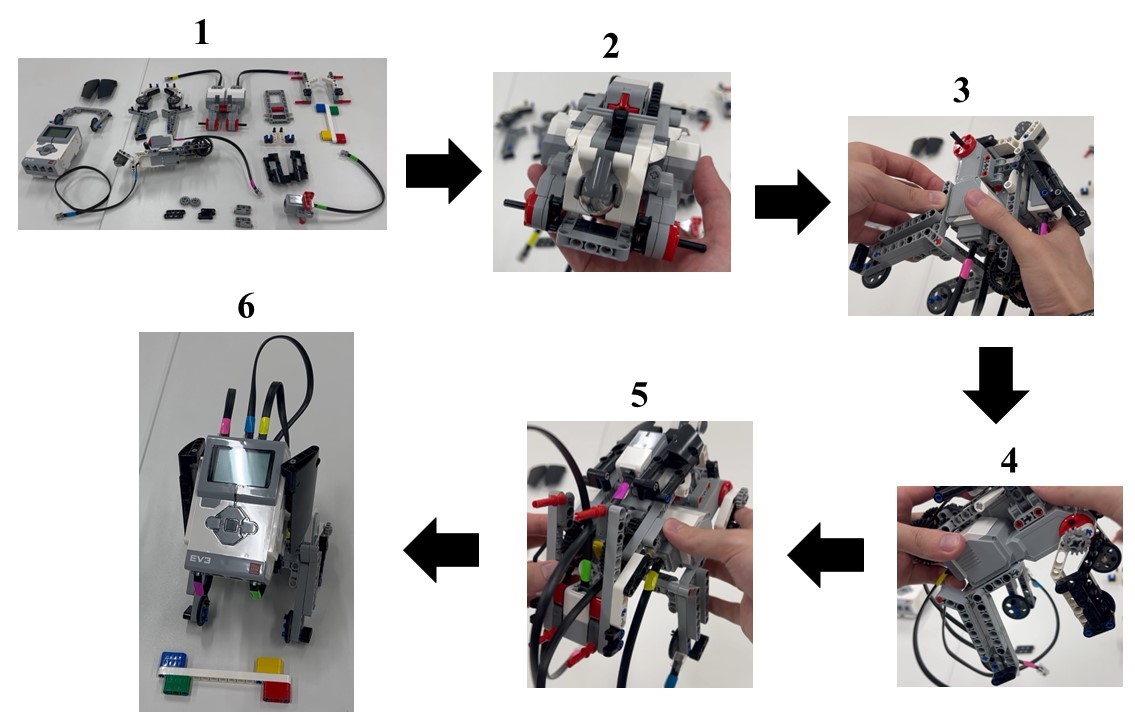}
\caption{Specific procedure for assembling experiment.}
\label{figa}
\end{figure*}

\subsection{Participants' relationships}
\begin{figure}[tbp]
\centering
\includegraphics[scale=0.27]{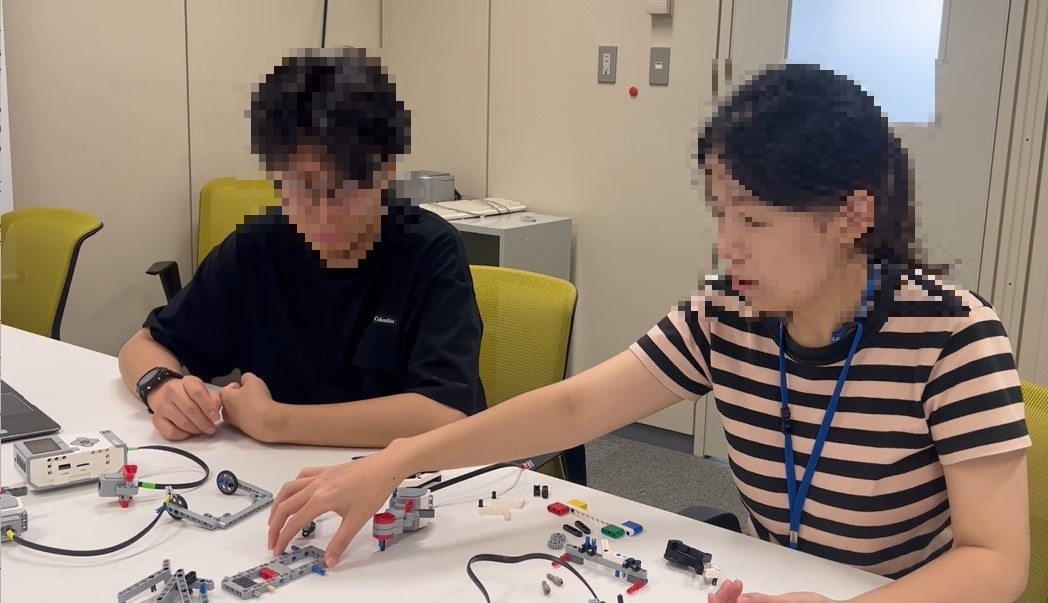}
\caption{Screenshots of agent and typing game during experiment.}
\label{fig4}
\end{figure}

\begin{figure}[tbp]
\centering
\includegraphics[scale=0.35]{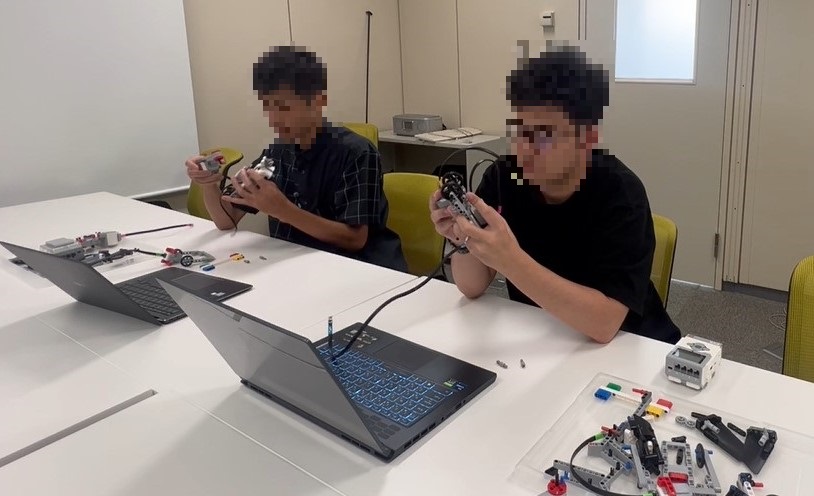}
\caption{Screenshots of agent and typing game during experiment.}
\label{fig5}
\end{figure}

Participants in this experiment performed the task in different relationships in the two conditions. 
The first was a condition in which the participants worked together to assemble the LEGO agent. 
In this condition, the two participants performed the task while watching the assembly video and checking with each other.
Although the relationship between the participants increased as they cooperated to perform the task, it is not clear how their empathy for the assembled agent changed. 
Figure~\ref{fig4} shows the cooperative condition.
\\ \indent
The second condition was for the two participants to assemble the LEGO agent by themselves.
In this condition, the two participants each performed a task while watching the assembly video. 
Although this was an individual assembly task, the reason why two participants performed the task at the same time was that there was a possibility that their empathy for the agent would change depending on the relationship between the participants in the same space. 
Figure~\ref{fig5} shows the individual assembly condition. 
In both conditions, each participant could watch the assembly video as they liked, and the participants were not restricted from talking to each other.

\subsection{Empathy object}
Two empathy objects were investigated in this study: empathy for the assembled LEGO agent and empathy for the participants who performed the task together. 
The reason for selecting the agent and another participant as the empathy target is that many previous studies have focused on only one of the two and have not compared the differences in empathy for the two through an actual task.
\\ \indent
We investigated the extent to which the promotion of empathy toward agents approximates the value of empathy toward humans, and whether the media equation of people treating artifacts like people affects empathy toward agents through the IKEA effect.

\subsection{Analysis method}
The analysis was a three-factor mixed design ANOVA. 
The between-participants factor consisted of two levels of participant relationships (cooperative and individual). 
The within-participant factor consisted of two levels before and after the IKEA effect and the empathy target (agent, another participant) empathy value.
\\ \indent
On the basis of the results of the participants' questionnaire responses, we investigated the influence of the IKEA effect and the participants' relationships as factors eliciting empathy for the agent and empathy for the other participant. 
The empathy values before and after the task were used as the dependent variable. 
R (ver. 4.1.0) was used for ANOVA.

\section{Results}
\begin{figure*}[tbp]
    \centering
    \includegraphics[scale=0.45]{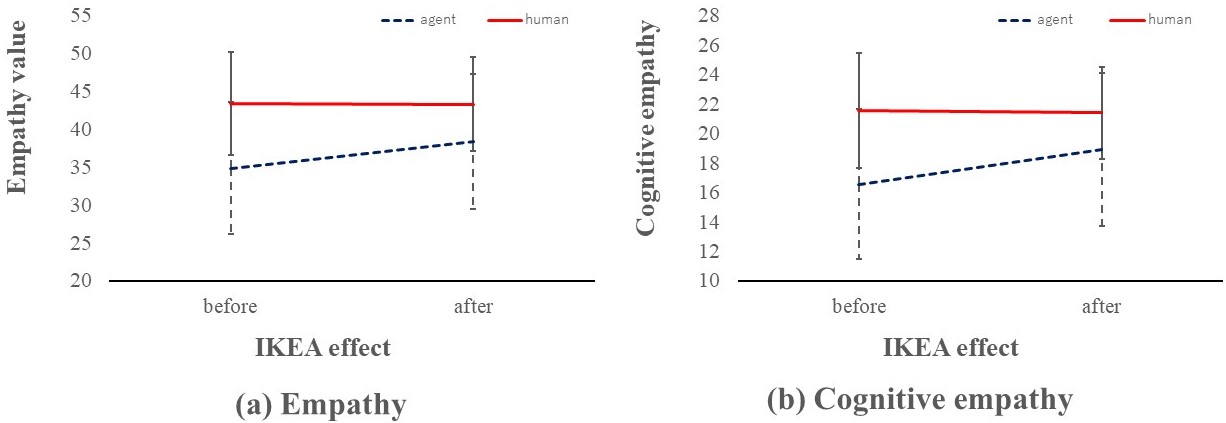}
    \caption{Interaction results for (a) empathy and (b) cognitive empathy.}
    \label{fig6}
\end{figure*}

ANOVA was used to analyze the responses to the questionnaire and classify empathy into the categories of affective empathy and cognitive empathy. 
Statistics for each empathy category are presented in Table~\ref{table2}. 
The raw data of the participants used in this study are available in Supplementary Information.
The results of the ANOVA are presented in Table~\ref{table3}, which shows that there was an interaction between the two factors, IKEA effect and empathy target. The results of an analysis of the interaction are shown in Figure~\ref{fig6}.
\\ \indent
There were no significant interactions between the participant relationship and the IKEA effect in either condition.
There were also no significant interactions between the participants' relationship and the empathy target. 
In the following, we omit a description of the main effects for items for which an interaction was found, and we present the main effect results for items for which no interaction was found and a main effect was found. 
Thus, since an interaction was found between the IKEA effect and the empathy target, the results of the analysis of simple main effects are presented in Table~\ref{table4}.

\begin{table*}[tbp]
    \caption{Participants' empathy statistical information}
    \centering
    \scalebox{0.85}{
        \begin{tabular}{c|c|c|cc|c|c|c|cc}\hline 
            \multicolumn{2}{c|}{Category} & Conditions & Mean & S.D. & \multicolumn{2}{c|}{Category} & Conditions & Mean & S.D. \\ \hline 
            & & cooperative-agent & 35.00 & 7.455 & & & cooperative-agent & 18.75 & 3.338 \\ 
            & & cooperative-human & 44.60 & 6.707 & & & cooperative-human & 22.60 & 3.560\\ 
            Empathy & before & individual-agent & 34.70 & 10.04 & Affective & before & individual-agent & 17.80 & 5.064 \\ 
            & & individual-human & 42.20 & 6.748 & empathy & & individual-human & 21.05 & 4.174 \\ \cmidrule{2-5}\cmidrule{7-10} 
            & & cooperative-agent & 38.95 & 9.185 & & & cooperative-agent & 20.10 & 3.986 \\ 
            (Q1-Q12) & & cooperative-human & 43.25 & 6.851 & (Q1-Q6) & & cooperative-humans & 22.20 & 4.491 \\ 
            & after & individual-agent & 37.85 & 8.744 & & after & individual-agent & 18.85 & 4.815 \\ 
            & & individual-human & 43.40 & 5.716 & & & individual-human & 21.60 & 3.952 \\ \hline
            & & cooperative-agent & 16.25 & 4.993 & \multicolumn{5}{c}{}  \\ 
            & & cooperative-human & 22.00 & 3.880 & \multicolumn{5}{c}{}  \\ 
            Cognitive & before & individual-agent & 16.90 & 5.360 & \multicolumn{5}{c}{}  \\ 
            empathy & & individual-human & 21.15 & 3.937 & \multicolumn{5}{c}{} \\ 
            \cmidrule{2-5}
            & & cooperative-agent & 18.85 & 5.697 & \multicolumn{5}{c}{}  \\ 
            (Q7-Q12) & & cooperative-human & 21.05 & 3.576 & \multicolumn{5}{c}{}  \\ 
            & after & individual-agent & 19.00 & 4.812 & \multicolumn{5}{c}{}  \\ 
            & & individual-human & 21.80 & 2.628 & \multicolumn{5}{c}{}  \\ 
            \cmidrule{1-5}
        \end{tabular}}
    \label{table2}
\end{table*}

\renewcommand{\arraystretch}{1.1}
\begin{table*}[tbp]
\caption{Analysis results of ANOVA}
\centering
\scalebox{0.9}{
\begin{tabular}{cllll}\hline
& \multicolumn{1}{c}{Factor} & \multicolumn{1}{c}{\em{F}} & \multicolumn{1}{c}{\em{p}} & \multicolumn{1}{c}{$\eta^2_p$}\\ \hline
& Participants' relationships & 0.266 & 0.609 \em{ns} & 0.007 \\ 
& IKEA effect & 5.451 & 0.025 * & 0.125\\
Empathy & Empathy object & 24.44 & 0.000 *** & 0.391 \\ 
(Q1-12)& Participants' relationships $\times$ IKEA effect & 0.346 & 0.560 \em{ns} & 0.009 \\ 
& Participants' relationships $\times$ Empathy object & 0.024 & 0.877 \em{ns} & 0.001 \\ 
& IKEA effect $\times$ Empathy object & 5.989 & 0.019 * & 0.136 \\
& Participants' relationships $\times$ IKEA effect $\times$ Empathy object & 1.279 & 0.265 \em{ns} & 0.033 \\ 
\hline
& Participants' relationships & 1.194 & 0.281 \em{ns} & 0.031 \\ 
& IKEA effect & 2.003 & 0.165 \em{ns} & 0.050\\
Affective & Empathy object & 20.17 & 0.000 *** & 0.347 \\ 
empathy & Participants' relationships $\times$ IKEA effect & 0.130 & 0.720 \em{ns} & 0.003 \\ 
(Q1-6)& Participants' relationships $\times$ Empathy object & 0.000 & 0.985 \em{ns} & 0.000 \\ 
& IKEA effect $\times$ Empathy object & 2.333 & 0.135 \em{ns} & 0.058 \\
& Participants' relationships $\times$ IKEA effect $\times$ Empathy object & 0.720 & 0.402 \em{ns} & 0.019 \\ 
\hline
& Participants' relationships & 0.034 & 0.854 \em{ns} & 0.001 \\ 
& IKEA effect & 6.686 & 0.014 * & 0.150\\
Cognitive & Empathy object & 19.85 & 0.000 *** & 0.343 \\ 
empathy & Participants' relationships $\times$ IKEA effect & 0.418 & 0.522 \em{ns} & 0.011 \\ 
(Q7-12)& Participants' relationships $\times$ Empathy object & 0.072 & 0.791 \em{ns} & 0.002 \\ 
& IKEA effect $\times$ Empathy object & 7.354 & 0.010 ** & 0.162 \\
& Participants' relationships $\times$ IKEA effect $\times$ Empathy object & 1.297 & 0.262 \em{ns} & 0.033 \\ 
\hline
%\multicolumn{5}{c}{}
\end{tabular}} \\ \hspace{-90mm}
            \em{p}:
{{*}p\textless\em{0.05}}
{{**}p\textless\em{0.01}}
{{***}p\textless\em{0.001}}
\label{table3}
\end{table*}
\renewcommand{\arraystretch}{1.0}

\renewcommand{\arraystretch}{1.1}
\begin{table*}[tbp]
\caption{Analysis results of simple main effect}
\centering
\scalebox{1.0}{
\begin{tabular}{cllll}\hline
& \multicolumn{1}{c}{Factor} & \multicolumn{1}{c}{\em{F}} & \multicolumn{1}{c}{\em{p}} & \multicolumn{1}{c}{$\eta^2_p$}\\ \hline
& IKEA effect in empathy for agent & 6.947 & 0.012 * & 0.155 \\ 
Empathy & IKEA effect in empathy for human & 0.014 & 0.905 \em{ns} & 0.000\\ 
(Q1-12) & Empathy object before IKEA & 29.07 & 0.000 *** & 0.434 \\
& Empathy object after IKEA & 10.55 & 0.002 ** & 0.217 \\ 
\hline
Cognitive & IKEA effect in empathy for agent & 9.447 & 0.004 ** & 0.199 \\ 
empathy & IKEA effect in empathy for human & 0.111 & 0.741 \em{ns} & 0.003\\ 
(Q7-12) & Empathy object before IKEA & 25.31 & 0.000 *** & 0.400 \\
& Empathy object after IKEA & 7.317 & 0.010 * & 0.162 \\ 
\hline
\end{tabular}} \\ \hspace{-60mm}
 \em{p}:
{{*}p\textless\em{0.05}}
{{**}p\textless\em{0.01}}
{{***}p\textless\em{0.001}}
\label{table4}
\end{table*}
\renewcommand{\arraystretch}{1.0}

\subsection{Empathy}
\begin{figure*}[tbp]
\includegraphics[scale=0.4]{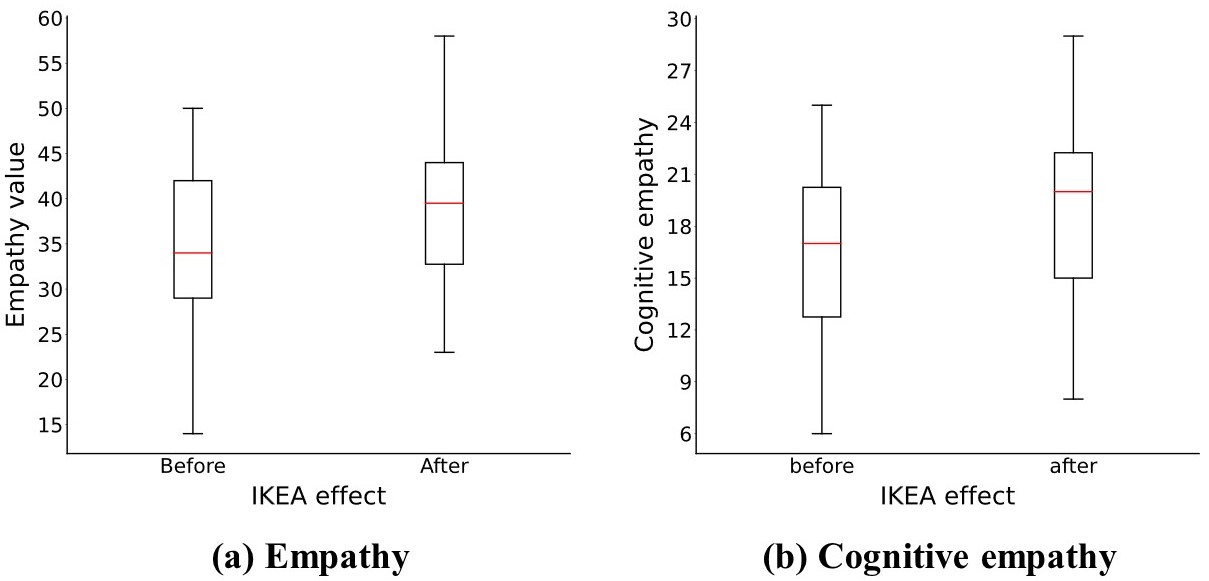}
\centering
\caption{Results of post-task (a) empathy value and (b) cognitive empathy represented by box plots. Red lines are medians.}
\label{fig7}
\end{figure*}

The empathy results (Q1-Q12) revealed an interaction between the IKEA effect and the factors of the empathy object. 
Although the main effects were also significant for both the IKEA effect and empathy object factors, the interaction between the IKEA effect and the empathy object factors is omitted because one was found.
\\ \indent
The simple main effect revealed a significant difference in the IKEA effect factor when the empathy object was the agent, as shown in Figure~\ref{fig7}(a) (before: mean = 34.85, S.D. = 8.728; after: mean = 38.40, S.D. = 8.869). 
Furthermore, significant differences were found in the simple main effects of the empathy object factors before and after the IKEA effect. 
These results suggest that the task generating the IKEA effect promotes empathy for the agent, whereas it has no significant effect on the participants' relationships.
\\ \indent
The results also indicated that initial empathy for those who participated in the experiment together was higher than empathy for the agent. 
This result was supported by the IKEA effect and the main effect of the empathy object.

\subsection{Affective empathy}
The results for affective empathy (Q1-Q6) showed that, unlike empathy, there was no interaction between the IKEA effect and the empathy object factor. 
Affective empathy was significant only for the main effect of the empathy object factor.
\\ \indent
In the LEGO agent assembly task, the factors in this study did not significantly promote affective empathy. 
On the other hand, as with empathy, there was a significant difference between affective empathy toward the agent and affective empathy toward another participant.
From these results, neither the IKEA effect nor the relationship between participants had a significant effect on affective empathy.

\subsection{Cognitive empathy}
The results for cognitive empathy (Q7-Q12) showed an interaction between the IKEA effect and the empathy object, as was the case for empathy. 
The main effects of both the IKEA effect and the empathy object were also significant but were omitted because of the interaction between the IKEA effect and the empathy object factor.
\\ \indent
The simple main effect revealed a significant difference in the IKEA effect factor when the empathy object was the agent, as shown in Figure~\ref{fig7}(b) (before: mean = 16.58, S.D. = 5.124; after: mean = 18.93, S.D. = 5.206).
Furthermore, significant differences were found in the simple main effects of the empathy object factors before and after the IKEA effect. 
These results suggest that the task generating the IKEA effect promotes cognitive empathy toward the agent, whereas it has no significant effect on the participants' relationships.
\\ \indent
The results also indicated that initial cognitive empathy toward those who participated in the experiment together was higher than cognitive empathy toward the agent. 
This result was supported by the IKEA effect and the main effect of the empathy object.

\section{Discussion}
\subsection{Supporting hypotheses}
The way to improve the relationship between humans and anthropomorphic agents is to have humans empathize with the agents.
This idea is supported by several previous studies~\cite{Gaesser13,Klimecki16}. 
Human empathy for agents is a necessary component for agents to be used in society. 
When agents are able to take an appropriate approach to human empathy, humans and agents can build a trusting relationship.
\\ \indent
In this study, an experiment was conducted to determine whether participants' empathy toward empathy objects (agents, people) changes before and after a task. 
We focused on the IKEA effect and participant relationships as factors that influence empathy. 
An experiment was conducted to investigate three factors, the IKEA effect, the participants' relationships, and the empathy object, and to investigate whether a LEGO agent assembly task could promote empathy. 
To this end, three hypotheses were formulated, and the data obtained from the experiment were analyzed.
\\ \indent
The results showed that empathy toward the LEGO agent was promoted after the IKEA effect occurred, which supported H1.
There was no effect on empathy for the other participant before and after the IKEA effect, only empathy for the agent was greatly promoted.
This result indicates that the IKEA effect, in addition to enhancing value judgments about artifacts, also significantly affects empathy toward artifacts.
\\ \indent
Second, H2 and H3, that empathy for the agent differs depending on the relationship of the participants in the assembly process, were not supported. 
Although studies with groups have shown that cooperative work improves the relationship between participants~\cite{Sun16, Fraune20, Pauw22}, in the present study, empathy for another participant did not change when working individually.
Similarly, there were no differences in empathy for the agent based on the relationship of the participants.

\subsection{Influence of IKEA effect and participant relationships on empathy agents}
As AI, robotics, and anthropomorphic agent technologies are used in society, this study focused on empathy toward agents to help people build better relationships with them. 
Although there have been studies on the IKEA effect to increase the value of artifacts, there have been no studies that investigated empathy for artifacts through the IKEA effect. 
This study shows that the IKEA effect is a factor that promotes empathy for artifacts. 
In particular, the IKEA effect was found to significantly promote cognitive empathy. 
It was revealed that when people assemble a LEGO agent, they acquire the agent's perspective and are more likely to fantasize about the agent.
\\ \indent
Although H2 and H3 were not supported, the IKEA effect in the assembly process promoted empathy toward the agent regardless of the relationship between the participants.
We can conclude that for people, the degree of assembly work and the opportunity to interact with the agent were not affected by the relationship between the participants in the assembly task.
\\ \indent
On the other hand, the results of this study revealed a significant difference between empathy for the agent and empathy for the person.
Empathy for the agent was not promoted to the same degree as empathy for the person. 
Although the media equation considers artifacts to be treated like people, different empathy judgments were made for the agent and person as empathy targets.
\\ \indent
The  finding of this research is that agents can actually be assembled to promote empathy toward them and improve relationships. 
This is a solution to the anxiety about agents that will permeate society in the future.
In addition, users' understanding of agents can be improved by assembling agents. 
As an example, by introducing an educational method using assembled agents in educational settings, effective education can be provided, and at the same time, empathy toward the agents can be promoted, which will lead to a better relationship and long-term use.
\\ \indent
In previous studies on empathy agents, Tsumura and Yamada investigated the influence of empathy on agent appearance and self-disclosure~\cite{Tsumura23-1}, task difficulty and task content, agent expression, and task success or failure~\cite{Tsumura23-2}. 
These results, when combined with the present study, will enable the design of empathic agents that are more easily accepted by people.

\subsection{Limitations}
One limitation of this study is the uniformity of the assembly time of the participants. 
Since this experiment had to be conducted in pairs, the time for the assembly process was not completely standardized.
Therefore, participants who completed the LEGO agent earlier may have had some influence on the interaction time with the agent and the relationship between the participants.
In future studies, we will investigate whether empathy for the agent affects the interaction with the LEGO agent after the assembly process.
\\ \indent
In addition, to simplify the assembly task in this study, the task was conducted with the small parts assembled in advance.
We also prepared a video of the assembly procedure so that the participants could visually understand it. 
However, in actual assembly work, it is conceivable that instructions would be given in a procedure manual instead of a video. 
If the assembly task becomes more difficult, it would be necessary to investigate whether the IKEA effect promotes empathy toward the agent.
\\ \indent
Other issues that need to be discussed include the level of difficulty of this experiment. 
Although this study showed that the IKEA effect affected empathy, it is possible that the difficulty level of the assembly task may have prevented participants from completing it and thus not affect their empathy toward the agent. 
Therefore, it is necessary to consider the level of difficulty and type of assembly tasks in the future.
\\ \indent
In addition, although we focused on the participants' relationships as a factor in this study, it is also necessary to consider their social relationships.
In this study, we did not specifically consider the participants' relationships in the recruitment process. 
However, when multiple people work in the same space during actual assembly, social relationships will be involved among people. 
Therefore, we will investigate whether social relationships influence empathy toward the agent in the future.

\section{Conclusion}
To solve the problem of agents not being accepted by humans, we hope that by encouraging humans to empathize with them, agents will be used more in human society in the future. 
This study is an example of how to promote empathy between humans and agents. 
The experiment was conducted with a three-factor mixed design, with the between-participants factor measuring the relationship between the participants and the within-participants factor measuring the change in empathy toward the target before and after the IKEA effect and with the empathy target. 
The results showed that there was no main effect for the participants' relationality factor and that after the IKEA effect, more empathy was promoted toward the agent by a statistically significant margin.
In addition, the participants' relationship factor did not show an effect on empathy toward the agent, and the degree of assembly work did not affect empathy. These results supported  one of our hypotheses.
This study is an important example of how human empathy works with artifacts. 
Future work will investigate whether the IKEA effect also promotes empathy for agents in online environments and will focus on the impact of interactions after the IKEA effect.

\section*{Data availability statements}
The datasets used in our study are available as additional supporting files.

\section*{Acknowledgments}
This work was partially supported by JST, CREST (JPMJCR21D4), Japan. This work was also supported by JST, the Establishment of University Fellowships towards the Creation of Science Technology Innovation, Grant Number JPMJFS2136.

%%===========================================================================================%%
%% If you are submitting to one of the Nature Portfolio journals, using the eJP submission   %%
%% system, please include the references within the manuscript file itself. You may do this  %%
%% by copying the reference list from your .bbl file, paste it into the main manuscript .tex %%
%% file, and delete the associated \verb+\bibliography+ commands.                            %%
%%===========================================================================================%%

\bibliography{sn-bibliography}% common bib file
%% if required, the content of .bbl file can be included here once bbl is generated
%\input sn-article.bbl

\end{document}